# Ballistic and snake photon imaging for locating optical endomicroscopy fibres

**M. G. T**ANNER,[1,2,*] **T. R. C**HOUDHARY,[2,3] **T. H. C**RAVEN,[2] **B. M**ILLS,[2] **M. B**RADLEY,[2,5] **R. K. H**ENDERSON,[4] **K. D**HALIWAL,[2] AND **R. R. T**HOMSON[1,2]

[1]*Scottish Universities Physics Alliance (SUPA), Institute of Photonics and Quantum Sciences, School of Engineering and Physical Sciences, Heriot-Watt University, Edinburgh EH14 4AS, UK*
[2]*EPSRC IRC Hub, MRC Centre for Inflammation Research, Queen's Medical Research Centre, University of Edinburgh, Edinburgh UK*
[3]*Institute of Biological Chemistry, Biophysics and Bioengineering, School of Engineering and Physical Sciences, Heriot-Watt University, Edinburgh EH14 4AS, UK.*
[4]*Institute for Integrated Micro and Nano Systems, School of Engineering, University of Edinburgh, Edinburgh EH9 3FF, UK.*
[5]*School of Chemistry, University of Edinburgh, Edinburgh, UK*
*\*M.Tanner@hw.ac.uk*

**Abstract:** We demonstrate determination of the location of the distal-end of a fibre-optic device deep in tissue through the imaging of ballistic and snake photons using a time resolved single-photon detector array. The fibre was imaged with centimetre resolution, within clinically relevant settings and models. This technique can overcome the limitations imposed by tissue scattering in optically determining the *in vivo* location of fibre-optic medical instruments.

## 1. Introduction

Image-guided minimally invasive interventions are increasingly being developed for the purposes of diagnosis and / or treatment, from optical endomicroscopy [1,2] or capsule endoscopy [3,4], to the use of robotically augmented surgical procedures. For many applications it is highly desirable to ascertain the precise localisation of the medical device, especially as target tissue size becomes smaller and further away from direct visualisation with the naked eye. Using light to facilitate the precise localisation of a medical device is already applied in many imaging techniques, for example, conventional endoscopes are guided to particular regions of the internal organs using a steering mechanism, an expert knowledge of anatomy and visual confirmation of correct placement. With the advent and increasing use of miniaturized optical endomicroscopy imaging [5] and sensing [6] inside the human body it is now possible to access anatomical regions traditionally out of reach of standard endoscopes. At present these fibre optic devices are often simply pushed out of the working channel of a scope (e.g. bronchoscope), or inserted without the use of any direct visual guide. In these scenarios, the final location of the distal tip of the optical endomicroscope is estimated from a knowledge of the position of the end of the scope, experience (by the clinician) and the nature of visible structures in the limited field of view of the fibre imaging system.

This paper describes a method of observing the location of the fibre optical endomicroscope using the small fraction of photons that escape the tissue with low scattering

– so-called "ballistic and snake" photons [7]. The high absorption and scattering of light in the visible spectrum has traditionally prevented exploitation of its use to provide information about tissue layers beyond a depth of a few millimetres under the skin. Fortunately, a near-infrared "optical window" [8,9] exists from ~ 700 to 900 nm, where the absorption coefficient is < 0.1 cm$^{-1}$. This window can be exploited to transmit light beyond several centimetres, but scattering as light bounces off the tissues in its path diffuses the light, reducing the spatial information transmitted from an embedded source (e.g. optical endomicroscope) to an external detector. However, a small number of photons can in principle travel directly from the source to a detector without ever being scattered *en route* [7,10,11]. Here we exploit these so called "ballistic" and slightly scattered "snake" photons to locate a 785 nm light source deep within biological tissues, with centimetre imaging resolution, by using time sensitive single-photon avalanche detector (SPAD) arrays.

Ballistic photon imaging has been commonly applied to extracting improved images of objects taken through scattering media, such as cloudy liquids and gases when back illuminated [7,10,11]. We apply a variant of this technique to successfully demonstrate the location of optical fibre probes (as an analogue of an optical endomicroscope) with centimetre imaging resolution in clinically relevant models. Our system is compact (a tripod mounted camera and a laser pulsed source) and can be applied within an environment with fluorescent room lights on (as in a clinical setting). The simplicity of this approach will enable ballistic photon localization to move out of the lab and into the clinic, permitting precise localisation / navigation in real-time, using inexpensive and compact equipment. We anticipate as the detector technology improves, that the resolution will improve to the sub centimetre regime.

## 2. Experiment

### 2.1 Principle of operation

The principle of operation is as follows (see Fig. 1): A time resolved single photon detecting imaging system is focussed on the tissue / body / organ (now referred to as 'the sample') and the optical fibre is placed within the sample. Short laser pulses are coupled into the proximal end of the fibre, which are then subsequently emitted from the fibre tip inside the sample as a large number of photons at time t = 0. A small number of ballistic and snake photons transit rapidly through the sample and are imaged at t = X, where X is defined by the speed of light through the sample and the distance from the fibre tip, giving accurate fibre location information. Weakly scattered photons (referred to as "snake" photons) arrive slightly delayed from the true ballistic photons, at t = X + $\Delta X_1$. These photons still provide useful spatial information about the fibre tip location and can be used in combination with ballistic photons to improve signal strength as required. Highly scattered photons arrive at t = X + $\Delta X_2$. These provide minimal spatial information (decreasing for higher values of $\Delta X_2$), and are ignored by the imaging system (or "time gated" out of the image). This is shown schematically in Fig. 1(d). The image obtained of the ballistic photons shows a source at the end of the optic fibre. "Snake" photons arriving later cloud around this point source, and further scattered photons will appear as an even more diffuse cloud. In experimental application, the transit time t = X does not need to be known as the observer is interested in distinguishing early arriving photons only relative to later scattered arrivals. In practice for the majority of demonstrations in this paper the fibre with defined Numerical Aperture (NA) was orientated perpendicular to the camera, as such almost all the detected photons have undergone at least one scattering event to be directed towards the camera and are therefore snake photons. Additionally truly ballistic photons are extremely improbable when the tissue thickness is more than a few mm [7,10,11], instead those snake photons that scatter but maintain a path close to the ballistic path are observed in this work. However, for conceptual clarity these will still be referred to as the ballistic photon arrivals.

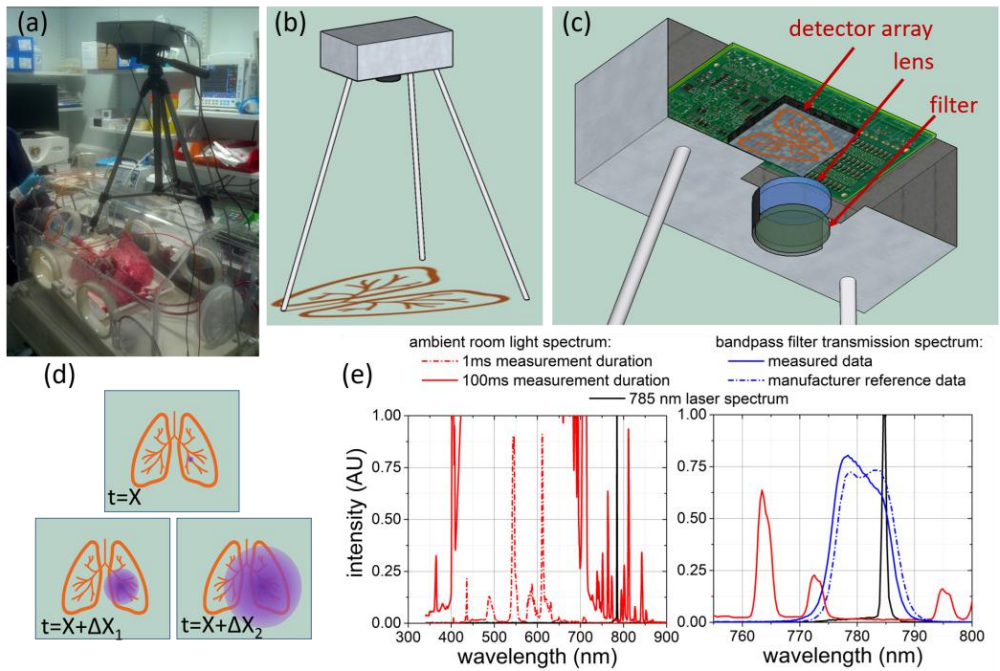

Fig. 1. Fibre probe locating camera prototype for use in ambient lighting conditions. (a) Camera positioned over ventilated and perfused ovine lungs under normal laboratory lighting conditions; (b) Schematic of camera arrangement over the object to be imaged; (c) Internal lens, filter and detector setup; (d) Schematic representation of the expected observed image at different photon arrival times (purple), with lung image overlaid (orange); (e) Spectra of room lights from standard fluorescent bulbs (solid and dashed red lines) shown at two different measurement durations to provide 100X scaling, near infrared bandpass filter transmission measured and manufacturer reference data (blue solid and dashed lines) and ~785 nm laser emission spectrum (black line).

## 2.2 Detector technology

To facilitate accurate imaging of the end of the fibre using the ballistic and snake photons, it was necessary to use a highly sensitive detector with high-timing resolution, such that the early-arriving ballistic and snake photons can be discriminated from the late-arriving highly scattered photons. To this end, we use an advanced 32 x 32 CMOS SPAD array known as "MegaFrame" [12–17] allowing timing of photon arrival in 50 ps bins with 200 ps jitter on all 1024 pixels simultaneously (further described in the Appendix), however the principle demonstrated here is not limited to any one detection technology. Importantly for our application, the two-dimensional nature of the MegaFrame array facilitates imaging when used with a simple camera lens as shown in Fig. 1(b) and (c). The ability to time the arrival of photons allows the disambiguation of ballistic and scattered photons and the arrangement of the detector system is describe further in the Appendix.

## 2.3 Photon arrival histogramming and image reconstruction

Time Correlated Single Photon Counting (TCSPC) is the well-established technique involving building histograms of photon arrival statistic over repetitive measurements [18]. While conceptually it is useful to consider the transit of a single pulse of laser illumination through a scattering sample (as in *Principle of Operation*), in practice the number of photons reaching the detector system is many orders of magnitude too small for useful measurement. Instead the laser is operated with a regular chain of pulses, and with the onset of each pulse each detector pixel has the opportunity to time the arrival of a photon. Over the course of many laser pulse repetitions the photon arrival times are collated in the form of histograms (as shown in later figures). It is important to note that such a technique is not a 'single-shot' measurement, but instead an ensemble test over the total measurement period. With the laser pulse repetition rates up to 80 MHz used here (see Appendix) the entire experiment was repeated every 12.5 ns, or performed $80 \times 10^6$ times in a 1 second measurement. The maximum number of photons per detector pixel counted is kept well below this number to avoid statistical error (known as photon pileup [18]) and detector saturation (see Appendix). The short interval between repetitions, and limited total measurement time, allows for the assumption that the system remains constant over the measurement period. Laser repetition rates are normally chosen such that a laser pulse has left the experimental system before the following pulse appears. Thus, the histograms constructed are an accurate representation of the measurement of a single laser pulse transiting the system within the time window (12.5 ns minimum used here).

As described, for the demonstrations in this paper TCSPC timing data is recorded for all image pixels, forming a data cube over many repetitions of the laser pulse, with resolution on the time axis of 50 ps, the Time to Digital Converter (TDC) bin size. While the detector array frame readout rate is slower than this (see Appendix), the TCSPC histogramming technique allows processed data to be reconstructed at this time interval. Plotting data from all pixels at a single time point of the collected histograms forms an image snapshot as shown in later figures, or sequencing such images one after another on the time axis forms a video of the progression of photon scattering as accompanying this paper (see Visualisations). Data for each pixel is also extracted along the time axis for more detailed observation of arrival delay induced by scattering, also included in subsequent figures. Processing of this data is further described in the Appendix, and interpretation is discussed in the context of Fig. 2 onwards.

*2.4 Operation with ambient lighting*

The experimental layout as shown in Fig. 1(a) is an approximation to the system being placed over a patient in clinic, while the biological models investigated included thick tissue samples to approximate *in-vivo* conditions (described in later sections and Appendix). The experimental system described here is intended as a practical method of determining optical endomicroscope device location during bronchoscopy (or other fibre optical endomicroscopy procedures). As such, realistic clinical environments had to be considered. Many TCSPC experiments are performed in a fully darkened lab, an unrealistic environment for this application. However, normal lighting in clinical environments is provided by fluorescent lighting which has a well-defined emission spectrum composed of discrete peaks concentrated in the visible range (Fig. 1(e)). Operating in the tissue "optical window" [8,9], the laser source at 785 nm is beyond the majority of visible room light emission, although small discrete peaks are still present, e.g. peaks at 772 and 795 nm are observed in Fig. 1(e), respectively 440 and 620 times smaller than the 540 nm visible wavelength peak of the fluorescent lights. Inclusion of an optical bandpass filter into the camera arrangement (see Appendix and blue curves in Fig. 1(e)) allowed only a small wavelength range of photons to reach the detector, encompassing the intended 785 nm photons but excluding almost all photons from fluorescent lighting (Fig. 1(e)). The filter pass band has partial overlap with the low amplitude fluorescent peak at 772 nm.

In addition to fluorescent room lights, a realistic clinical environment also includes a multitude of monitors and equipment emitting low levels of background light. These will have minimal emission in the range used in this system (outside of the visible spectrum), but some overlap must be assumed. The nature of a TCSPC experiment is that photon detection is synchronised with the pulsing of the source laser. Detected light that is not synchronised will contribute a constant background on a resulting TCSPC histogram, rather than altering the shape of photon detection features from the synchronised source. Addressing background light can therefore be seen as a 2-fold approach, aggressive optical filtering to remove the vast majority of unwanted photons (to a level where they do not dominate photon detection statistics) and subtracting remaining unwanted photon detection with the dark counts (described further in the Appendix) as they do not contribute to the timing features in the photon detection histograms. This approach enables the system to be operated in a laboratory with main fluorescent room lights and numerous equipment monitors on with no apparent detriment to the fibre location determination. Fig. 1(a) shows actual lighting conditions for the data shown later in Fig. 5.

## 3. Results

Further details of the models used, experimental arrangements and acquisition parameters of all the following data are included in the Appendix for reference. All accompanying videos are compiled in Visualisation 1, or separately as referenced in the figure captions and detailed in the Appendix.

### 3.1 Observation of scattered light-in-flight

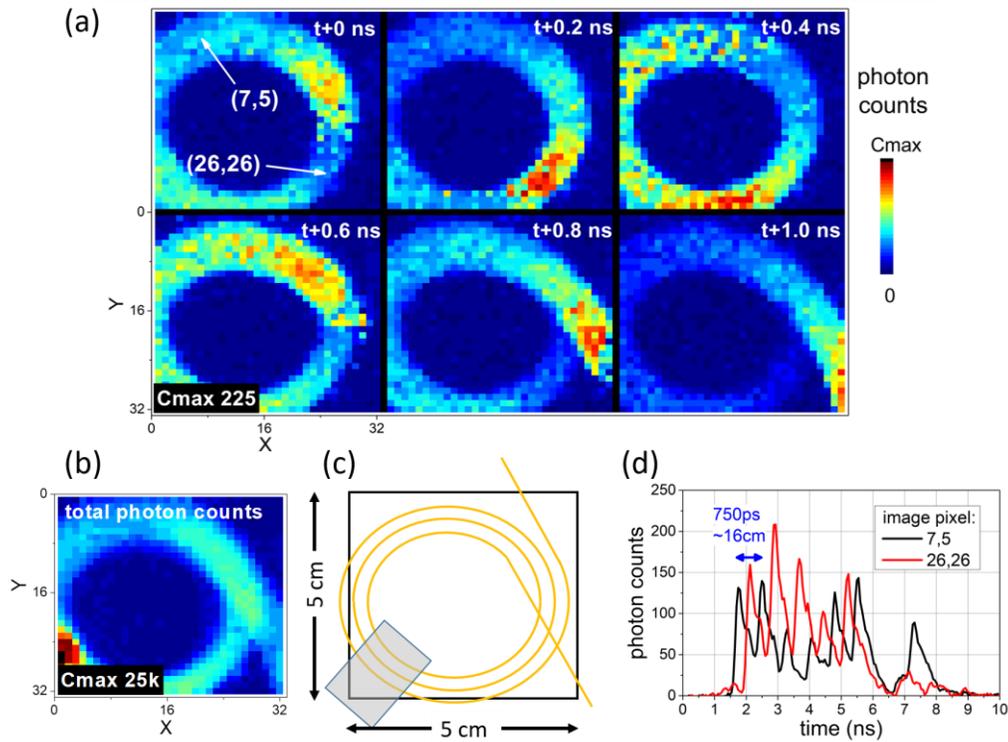

Fig. 2. Observation of 'light-in-flight' along an optical fibre, demonstrating function of the camera. (a) Pulses of light travelling around a coil of fibre shown at 200 ps intervals. Pulses are travelling in a clockwise direction, observed to chase each other around the coil of fibre. Image is deliberately defocussed to enable easier viewing of the fibre coil; (b) Sum over time of the same data; (c) Schematic of fibre coil; (d) TCSPC traces from individual pixels labelled. Video accompanying this figure offers clearer visualisation than the individual frames (see Visualisation 2). In this example, transparent tape was used to fix the coil of fibre in position at the bottom left of the image, causing increased photon scattering in this region most easily observed as the brightest region in (b).

The time resolved imaging modality exploited here ("light-in-flight" imaging [15,19]) was demonstrated with a coil of optical fibre (Fig. 2(c)) to aid understanding of image reconstruction and later results. A pulse of 785 nm light travelling along an optical fibre undergoes scattering, a limited number of the scattered photons are detected by the camera and can be used to observe the passage of the pulse along the fibre length [20]. A series of images were reconstructed such that the time axis of the data becomes the time axis of a video with photon count intensity as the z-axis. Frames of this are shown in Fig. 2(a), while the total photon counts (a sum over all time intervals) is shown in Fig. 2(b). *Cmax* is used in figures from herein to label the maximum of the *photon counts* colour scale on image plots. In the above example a single pulse of light is observed to circulate a coil of fibre 5 times (see Visualisation 2). Data from individual pixels shown in Fig. 2(d) demonstrates that the photon flux varies periodically on each pixel as the light pulse circulates. In Fig. 2(d) and later figures, reference to the image pixels (e.g. of Fig. 2(a) and (b)) is made with coordinates in the form (x,y). While the image frames of Fig. 2, or the accompanying Visualisation, most accessibly show pulse propagation it is important to examine the timing histogram Fig. 2(d) for full interpretation. A peak occurs in each data trace at the moment on the time axis when a camera pixel is bright. Peak amplitude varies as the fibre passes on top or underneath the previous coils. The shape of this peak represents a convolution of the laser pulse shape and the detector Instrument Response Function (IRF). Each data trace has peaks repeating at ~700 ps, the time taken for light to propagate a complete circuit around the fibre coil (~16 cm circumference, at the speed of light in fibre). There is a ~350 ps (~8 cm transit length in optical fibre) timing lag between the chosen pixels, equating to transit round approximately half the coil of diameter ~5 cm. As such the two traces are out of phase. In subsequent demonstrations there is no repeated observation of the laser pulse as it propagates only once through the sample. Instead the histograms are used to observe where the light first emerges from the sample, and the delay in light emerging from other regions. As such it is the leading edge of the IRF that is the key feature in such histograms, unaffected by the tail shape of the IRF observed in Fig. 2(d).

To demonstrate the application of this camera to imaging photons scattered in media, an optic fibre was suspended in a tank of an emulsified colloid (semi-skimmed milk, 1.5% butterfat), the uniformity of the medium providing a simple demonstration (Fig. 3). The fibre was orientated to point perpendicular to the axis of the camera, such that only light undergoing at least one scattering event would be detected.

"Lesser-scattered" light (ballistic and snake photons) reaches the imaging detector at an early arrival time, t~ 0.5 ns, highlighted on the top frame of Fig. 3(a) at pixel (22,15) - referring to the pixel at coordinates (x,y) on the labelled axis. More scattered light arrives later as expected (with delays of the order of nanoseconds) in the middle frames of Fig. 3(a) (t~3 ns and t~6 ns). For comparison, an image of the total photon counts, without time discrimination, is shown (Fig. 3(a) bottom frame) in which all data along the time axis has been summed. Comparing photon arrival statistics across the field of view in Fig. 3(d) shows delays of the order of 1 ns, equivalent to an optical distance of ~20 cm in a liquid with the refractive index of water. The total field of view in this experiment was 15 cm, suggesting photons underwent a sequence of complex scattering events to significantly increase the travelled optical path length before reaching the detector at ~ns delays.

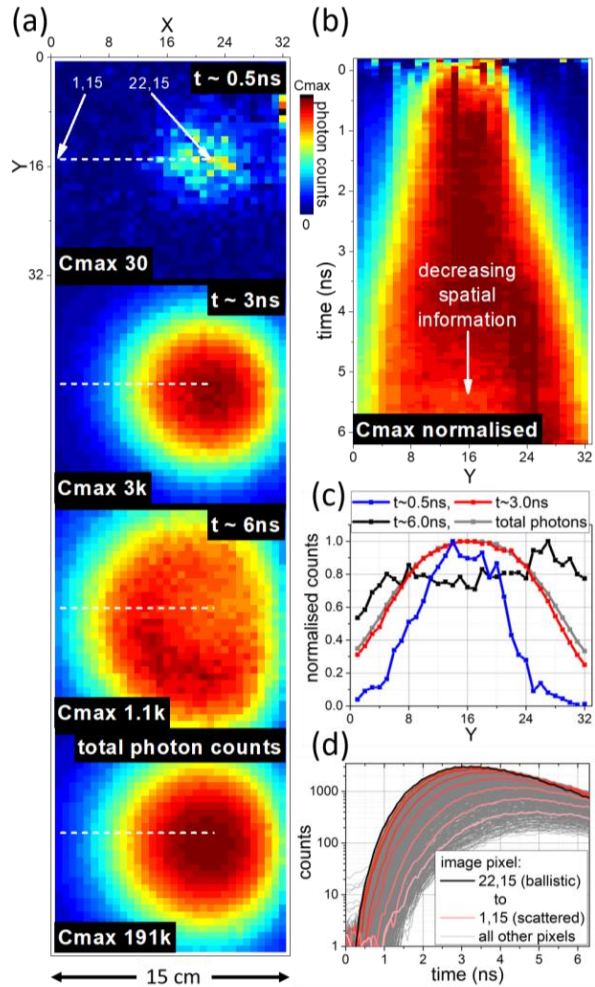

Fig. 3. Time resolved images of light scattered through a colloidal suspension (milk). (a) Time resolved photon arrival at t ~ 0 ns, t ~ 3 ns, t ~ 6ns, and the total photon counts over all time intervals (see Visualisation 3). Early arriving "ballistic" photons are observed in the top image; (b) & (c) Normalised profiles through the images showing the extent of the scattered spot of light with time in (b) and at various time points in (c); (d) Histograms of photon counting statistics for pixels at the centre of spot (22,15 – black line), moving to the edge of the spot (1,15 – pink line), and for all other pixels in the image (grey).

In the frames shown in Fig. 3(a) the scattered photon cloud is observed to progress towards the bottom left of the image, in the direction light was launched from the fibre optical probe. At t ~ 6 ns, a 'scattering wavefront' is in effect being imaged, with notably less light behind the direction of propagation of the original light pulse. The non-time resolved image (bottom of Fig. 3(a)) shows scattered photon detection concentrated in a circular pattern, giving some indication that the optical fibre is located centrally in this region. However, pulse propagation has moved the centre of this scattered photon cloud away from the true location of the optical fibre emission. Fig. 3(b) & (c) demonstrate how the cross-section of the scattered photon arrivals increases as time progresses, decreasing available spatial information. It can again be observed that the centre of the photon cloud progresses across the plot (to higher Y values) as the pulse propagates.

Moving in 3 pixels (~1.5 cm) increments from the centre of ballistic and snake photon arrival (pixel coordinates 22,15) along the dashed line shown in Fig. 3(a) there is little initial change in overall illumination intensity. Meanwhile in the photon arrival time histograms (coloured lines in Fig. 3(d)) there is a clear (if slight) change in photon arrival time statistics. Critically it is these photon arrival histograms (Fig. 3(d)) that demonstrate the early arrival of the least scattered photons (black line) before those travelling longer scattering paths (red shaded lines as examples, all other pixels in grey), while the associated images offer more convenient visualisation.

*3.2 Fibre location in tissue models*

The potential interference from significant muscle and bone structure was investigated in this technique using a number of models. This included an avian tissue model (whole chicken, plucked and cavity emptied, ~25 cm thickness), with the optical fibre inserted into the cavity and an explanted sheep lung (~15 cm thickness) placed behind a section of rib cage, thickness ~ 10 cm, (with the optical fibre inserted along a bronchi and pushed into an unknown location in the distal lung). In both cases the optic fibre was believed to be perpendicular to the axis of the camera based on insertion angle, precise location was unknown in advance due to lack of visibility.

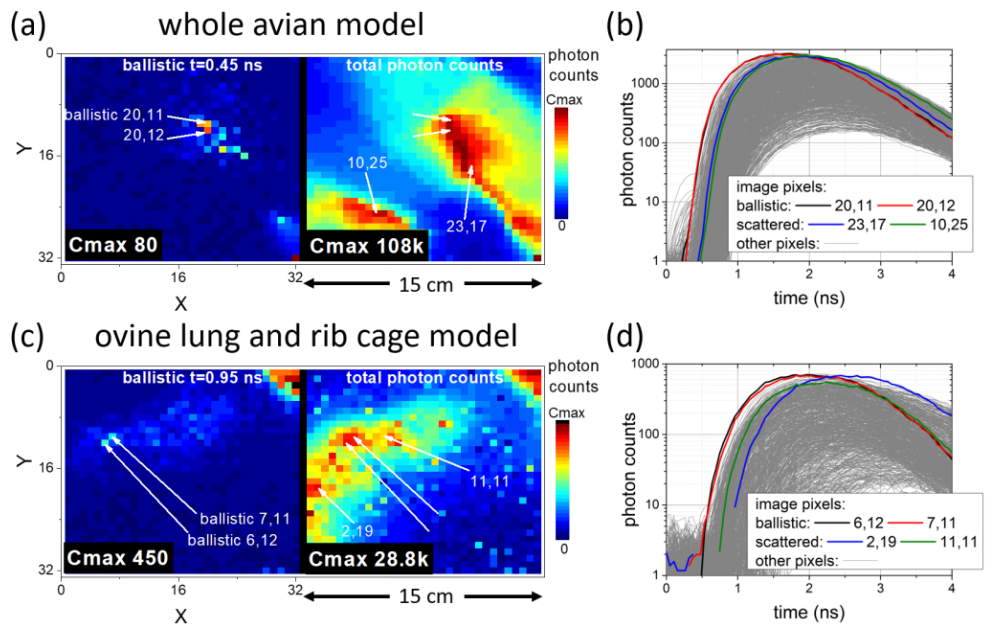

Fig. 4. Fibre location in tissue models. (a) A whole avian (chicken) model; (c) An ovine lung and rib cage. (a) & (c) Early ballistic photon arrival image locates the end of the fibre, and the total photon image (sum over all arrival times) offers little location information (see [Visualisation 4](#)); (b) & (d) TCSPC histograms confirm early arrival of ballistic photons and later arrival from other bright features in the image. (d) Excludes the data from the top right portion of the image, where light escaping from the fibre without passing through tissue was directly observed.

The line plots in Fig. 4(b) & (d) show the photon arrival statistics (TCSPC histograms) from all pixels, highlighted for the early arriving ballistic photons and other labelled bright points chosen from the non-time-resolved images as labelled on Fig. 4(a) & (c). Early arriving photons are notably ahead of all other pixels and easily identified (black and red lines in Fig.

4(b) & (d)), ~0.2 ns ahead of other bright regions in the image (green and blue lines in Fig. 4(b) & (d)). Visualisation of this on the image frame at an early arrival time ("ballistic" images) in Fig. 4(a) & (c) shows the fibre source as notably brighter than other pixels. Meanwhile the non-time resolved, total photon counts, image (sum of all time resolved image frames) instead shows evidence of the complex tissue and bone structure. Fig. 4(a) has other semi-bright pixels in the ballistic arrival image where photons are observed with moderate intensity at an early arrival time, however the timing histograms allow disambiguation of the earliest arriving photons. The source of these features is discussed further after Fig. 5.

### 3.3 Bronchoscopy of ventilated ovine lungs

Demonstration of the practicality of the method described here was performed by mounting the imaging camera over a size relevant biological model, comparable to placing the unit over a patient. Lighting in the room was kept on to emulate a clinical environment (room lights were turned off for comparative measurements). Additionally, a notable amount of broadband background light was always present from monitors and equipment.

A whole ovine lung (without ribcage, ~15 to 20 cm thickness) was ventilated and perfused, as has been described elsewhere [6]. An optical fibre probe was introduced into the lung through the working channel of the bronchoscope and into specific targeted segments. However, as is well known, a bronchoscope operator has limited knowledge of the location of an optical endomicroscope once it is pushed beyond the bronchoscope tip.

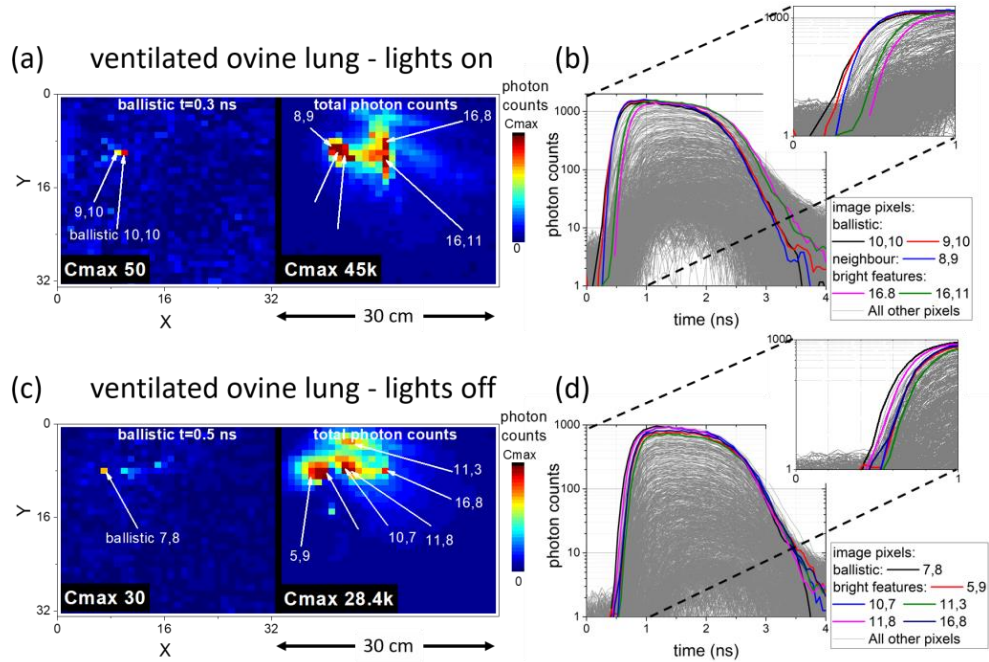

Fig. 5. Fibre location in a ventilated and perfused ovine lung model with the camera on a tripod over the 'patient'. (a) & (b) in the light and (c) & (d) in the 'dark'. The fibre was repositioned to a neighbouring segment between experiments. (a) & (c) Early ballistic photon arrival images locate the end of the fibre in contrast to total photon images (sum over all arrival times) offering little location information (see Visualisation 5). (b) & (d) TCSPC histograms confirm the ballistic arrival as compared to other bright features in the total photon image. Insets shown on shorter time scale for clarity of timing differences.

Fig. 5(a) & (c) again show ballistic photons arriving well localised to 1 or 2 pixels (~1 cm per pixel) as opposed to the non-time resolved images which show structure representative of the complex tissue scattering. Between experiments the fibre was moved to a neighbouring lung segment, the ballistic imaging shows a movement of ~3.5 cm. Meanwhile non time-resolved imaging shows a complex tissue structure in both cases, with no clarity as to fibre location. With room lights on, a greater 'dark' count rate was subtracted from Fig. 5(b) (28 counts per TDC bin, or 9.3 kHz) compared to (d) (2.2 counts per TDC bin, or 1.2 kHz), but not significant compared to photon counts and not detrimental to fibre location ability.

In the "ballistic" images for the tissue examples shown (Fig. 4 and Fig. 5), pixels of the image other than the labelled ballistic arrival are also brighter than the background level, such as the two bright pixels in the left image of Fig. 5(c). This is likely caused by a preferential scattering path, providing a low attenuation route for photons to escape from the tissue, perhaps relating to airways in the lung tissue in this case. Here, even though some increased scattering has occurred, slightly delaying the detected photons, a low attenuation path causes a relatively high signal in such an image. Examination of the timing properties in the line plots in Fig. 5(d) clearly distinguishes these pixels from the true fibre location, which shows earlier photon arrival. Such observation highlights the importance of using the timing statistics, not solely signal amplitude at specific time points, to distinguish the least scattered photon arrivals.

*3.4 Fibre location through dense tissue / human body*

While ballistic/snake photon emission through ovine lung and rib cage is believed to be a reasonable approximation to the structure of a human chest, transmission through entire human body sections was trialled to determine the limits of the current prototype with the optical fibre placed in a mount that could be positioned flush against the tissue surface, orientating the fibre facet in the direction of the camera. Imaging was performed through the palm of a hand (~4 cm total thickness) and human torso, just below the ribs imaged from the front of the body with the fibre placed against the back (~25 cm total thickness, ~95 cm torso circumference). Acquisition times were 17 seconds total exposure.

Numbers of photons exiting through the palm of a human hand (Fig. 6(a)) show clear ballistic photon arrival centred on a single pixel (~ 1 cm) in contrast to the image of the total scattered photons exiting the tissue from a large region of the hand. Some saturation of the non-time resolved data is observed as settings were kept constant between this and the more challenging transmission through human torso, but this does not affect the timing properties and therefore the ability to locate the fibre.

Transmission of NIR photons through the human torso with the fibre placed on the back, below the rib cage, was possible, although weak by comparison to the human hand. Ballistic/snake photon arrival is observed as shown in Fig. 6(c) to be stronger than photon counts in surrounding regions and well located (~1 cm resolution). To ease disambiguation from noise, images from three sequential time points are shown in which the same pixel location shows stronger photon arrival (see also [Visualisation 6](Visualisation 6)). Furthermore, the histogram (Fig. 6(d)) shows the ballistic photon arrival in the lead of all other pixels. Data for the pixels to the right of the ballistic arrival are highlighted (coloured lines in Fig. 6(d)), as these have the greatest total number of photons arriving (see total photons exiting tissue in Fig. 6(c)) but clear timing lag behind the ballistic arrival. Timing delay is once more seen to increase for pixels spatially further away from the ballistic arrival location, and these pixels have higher counts at increased delay (than the lead pixel) causing greater total counts in the non-time resolved plot.

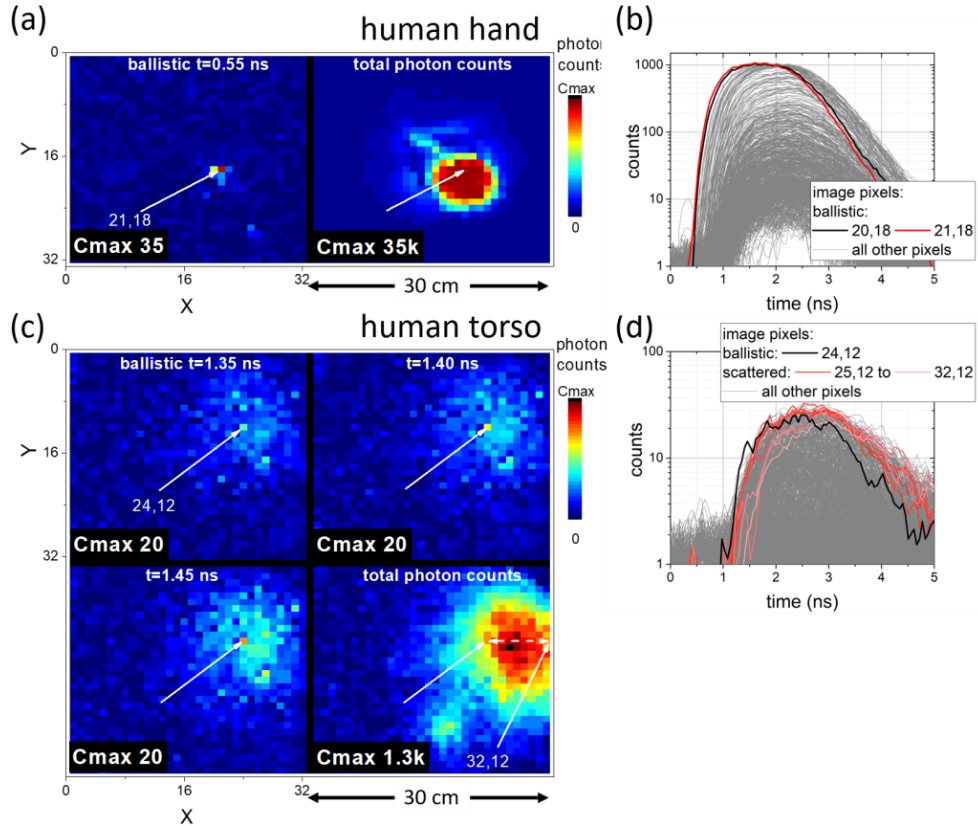

Fig. 6. Fibre location through the human body. (a) & (b) Human hand and (c) & (d) Human torso, below ribs, imaged from the front of body with the fibre placed against back. (a) & (c) Location of the optical fibre probe through early ballistic photon arrival images, in contrast to total photon images (sum over all arrival times). In (c) three subsequent time resolved frames are shown to aid in determination of ballistic arrival bright point above the noise level (see Visualisation 6). (b) & (d) TCSPC histograms demonstrating differences in photon arrival statistics, even in low photon count data for human torso.

Successful demonstration of this prototype in this scenario is challenging, as the light must pass through latissimus dorsi muscles, abdominal visceral organs and muscles of the abdominal wall. While signal to noise is reduced the approach was still successful. Background lighting was limited to reduce 'dark' counts (3.0 counts per TDC bin, or 40 Hz, in Fig. 6(c) & (d)). As shown here this result is close to the limits before detector noise makes it impractical to observe ballistic arrival in the form of the simple images presented, improvements in signal to noise are described in the following section. However, the greater scattering in this scenario leads to longer delays in photon arrival (Fig. 6(d), up to 2 ns rise times) which with sufficient numbers of photons counted eases the distinguishing of the ballistic from the scattered photon arrivals when photon arrival statistics are examined.

## 4. Discussion and future work

This work clearly illustrates that ballistic photon imaging can be used to locate an optical fibre *in vivo*. The success of this technique depends critically on counting sufficient photons exiting the tissue to provide good photon statistics. Unsurprisingly, as larger or denser (with a reduced photon mean free path) biological models are used, greater photon scattering is introduced, adding greater delay to the most scattered photons and offering a clearer demonstration of the advantages of a time resolved approach. However, this also reduces the number of ballistic (or near ballistic) photons exiting the tissue.

Photon flux reaching the camera could be improved by integrating an optic / scattering element to the fibre tip, as currently light exits within the NA of the fibre probe and not necessarily in the direction of the camera. Indeed, if the probe is in a non-scattering cavity, the current arrangement may give inaccurate location information as the photons propagate a short distance before undergoing a first scattering event towards the camera. If laser power is emitted in an omnidirectional manner it may be safe to increase power levels / repetition rates. We operate in a regime limited by the power of the available laser source (~7 mW at 80 MHz repetition rate). Higher powered lasers would offer improved signal at short measurement times, but safety would limit application in tissue.

If the approach here were to be applied to thicker/denser tissue structures, photon capture may also need to be improved to achieve practical measurement durations, however in this demonstration significant inefficiencies were present due to the use of a simple small aperture lens and a low fill factor (~1%) detector array. Detector arrays similar to that used here have now been developed with 45% fill factor, improved detection efficiency in the NIR and USB3.0 frame readout rates [21]. The fill factor improvement alone would reduce the 17 s exposure time through a human torso to ~ 3Hz image update rate. Improvement here could provide real-time imaging of medical device location, and enable plotting of the path travelled during procedures.

Choice of wavelength is key to both reduce absorption in the tissue, and avoid overlap with background light sources (e.g. room lights). The work here has demonstrated that NIR (in this case 785 nm) is a reasonable balance of these factors, key to enabling transfer to a normally lit environment. Other wavelengths in this region can be investigated in the future for comparison of the tissue scattering properties. While filtering of background light could be tightened around the pump wavelength, this was not a limiting factor in these initial demonstrations, but would enhance signal to noise (from ambient lighting) for demonstration through dense tissue with normal lighting conditions.

It is notable that the work here takes advantage of timing information on a finer scale than the intrinsic jitter of the SPAD detectors (~200 ps). The response characteristics of the SPAD pixels are defined by a repeatable Instrument Response Function (IRF), which is convoluted with the true timing characteristics of the photon arrivals. In this sense, finer timing information is not lost, and as performed here comparing timing statistics from pixels with comparable IRFs allows extraction of information not limited to this intrinsic figure of merit for timing accuracy. However, use of detectors with improved timing accuracy may indeed offer more obviously apparent differences between photons with different scattering paths.

As noted in this discussion, accurate location of the fibre tip is limited by a complex combination of factors. Signal to noise or photon budget (linked to tissue photon transmission statistics, detection efficiency and background light filtering) limits the frame rate of a realistic imaging system in these demonstrations to multiple seconds, but with improvement possible from greater photon capture. Results here were not limited by detector timing accuracy (in combination with the extent of tissue scattering delays in realistic samples). However complex tissue samples with cavities or varying refractive index structures may distort the imaging. It has not been possible to ascertain the 'ground truth' of fibre location deep inside the models. Future work will include the use of fiducial markers within complex tissue phantoms, allowing correlation of the location of a fibre probe determined through time resolved imaging to a ground truth (obtained through dissecting the phantom).

This technology has not yet demonstrated the ~mm accuracy observed with electromagnetic systems for detection of wired coils designed for integration with medical devices [22]. However, compatibility with existing optical endomicroscopy fibres motivates future investigation of the limits of accuracy of this principle. Meanwhile observed ~cm imaging resolution is for instance appropriate for device location during clinical optical endomicroscopy investigations of different lung segments.

Similar principles can be extended to fibre length location with photons scattered from the length of an optic fibre, however in a conventional optic fibre or optical endomicroscope this signal will be many orders of magnitude weaker than the light emitted from the fibre end. A deliberately highly scattering fibre optic could be employed to light up the device path, and prior knowledge that the source is in the form of a line with gentle bends will allow advanced image processing techniques to be employed to reconstruct accurate images of the location of the fibre path. This would allow the clinician, for example, to observe the path of a particular bronchi/bronchiole in the case of lung investigations. In clinical scenarios it is conceivable to perform imaging from two orthogonal axis, to enable three dimensional localisation, while overlaying the result on conventional imaging data for co-localisation.

A variety of related low photon number imaging techniques use advanced computation to improve image quality [20,23–25]. Expected temporal correlations between photon arrivals at neighbouring image pixels can be exploited to improve image quality [23] or overcome detector array timing limitations [24]. Recently a technique was demonstrated [25] to computationally exploit later arriving scattered photons in addition to ballistic photons to reconstruct images through 15 mm thick tissue phantoms. Application of these techniques to this work would be of great interest. However it is notable that the ideal nature of the bright point source emission from the fibre probes allows successful ballistic photon imaging of the location without complex computation, easing the transition to clinical application.

The method described here is highly compatible with existing fibre optical endomicroscopy technologies. The only required modification to any existing imaging or sensing system is the ability to couple short pulses of laser light into the fibre at the proximal end. The fibre location system consists of inexpensive and compact components and can be added easily to existing clinical applications without infrastructure changes. With the advent of minimally invasive surgery and robotic augmentation, these approaches have significant clinical utility.

## 5. Conclusions

We performed location of an optical fibre probe as an analogue of an optical endomicroscope through ballistic photon imaging with centimetre imaging resolution. The light escaping from complex tissue in a non-time resolved image does not give clear location of the fibre tip. Instead the image represents the least absorbing routes for the photons to exit the specimen. Meanwhile ballistic photon imaging gives very clear fibre tip location to within a centimetre, localised to a single pixel or neighbouring pixels in images, suggesting these are indeed very close to true ballistic photons having undergone minimal scattering. Realistic models of a chest and lung as well as a ventilated lung through a bronchoscope in an environment similar to that of a clinical setting have been investigated. Demonstration of the technique through the entire thickness of a human torso was successful. The system is compact and practical, even in rooms with fluorescent light illumination. This technology has potential for impact in medical procedures.

**Code availability**

The code is available via Edinburgh DataShare (http://dx.doi.org/10.7488/ds/2108).

**Data availability**

The experimental data is available via Edinburgh DataShare (http://dx.doi.org/10.7488/ds/2108).

**Appendix**

*5.1 Camera arrangement*

The MegaFrame was placed inside a metal box to minimise ambient light. A simple aspheric lens (focal length 4.6mm, NA 0.53, Thorlabs A390TM-B) was placed in the front of the box, which imaged a ~30 cm field-of-view, 50 cm from the camera, onto the 1.6 mm square detector array. Controlling working distance (lens to detector spacing) allowed adjustment of field of view (in conjunction with the distance to the object) to 15 cm for data in Fig. 3 & Fig. 4. A 5 cm field of view (lens focal length 13.86 mm, NA 0.18, Thorlabs C560TME-B) was used in Fig. 2.

A filter is placed in front of the lens of the camera (Thorlabs FL780-10, optical density >5 in visible range and >3 in the near infra-red) to reduce the influence of room lighting as seen in Fig. 1. All spectral data was collected with a compact spectrometer (Ocean optics USB2000+), except filter reference data from the manufacturer. Experimental filter transmission data shows some distortion due to the non-uniform spectrum of the broadband illumination source used in this measurement, reference data is shown for clarity (blue dashed line in Fig. 1(e)).

The detector array has a ~1% fill factor, which is the result of the gaps between the individual detectors of size ~ 6 µm which are placed on a ~ 50 µm pitch. Each SPAD on the Megaframe has independent Time Correlated Single Photon Counting (TCSPC) capability with a ~50 ps timing resolution (defined as the duration of distinct timing bins in the TCSPC histograms for each pixel). The timing jitter of the SPAD detectors is closer to 200 ps, defined as the Full Width Half Maximum (FWHM) of the TCSPC histogram generated from a pulse of light with duration much less than the detector timing jitter. This defines the uncertainty with which a feature in a TCSPC histogram can be assigned an accurate arrival time. However, in the experimental modality described here the 50 ps timing resolution describing the accuracy of the photon detector array TDC is the more relevant figure, while the jitter can be expected to cause some blurring/smoothing of features in time (a convolution of the true photon arrival statistics and the instrument response).

*5.2 Illumination*

The light source for our demonstration was a ~785 nm laser diode (Picoquant LDH-D-F-N-780), with combined pulse duration and timing jitter (FWHM) as low as 80 ps at 50 % maximum power (tested by manufacturer), notably increasing to ~1 ns at maximum power due to an asymmetric tail in the instrument response function. However, the leading edge of the laser pulse remains sharp (rise time <40 ps) in all scenarios. As this work is examining the early arrival of ballistic and snake photons, it is the sharp leading edge of the laser pulse that is primarily required. The laser pulse repetition rate was varied between 20 MHz and 80 MHz, at a range of powers up to 7.4 mW average output. In the optimised system with operation in ambient lighting conditions, high laser power was chosen to minimise exposure time (see Table 1) to increase the signal (laser light exiting tissue) to background (ambient room lights and dark counts) in each exposure.

*5.3 Detector operating regime*

The detector system can read out TCSPC data at a rate <2000 counts per second per pixel (or frame rate), limited by the USB2.0 interface. Later iterations of the detector system can achieve ~2 orders of magnitude higher rates through a USB3.0 interface. The detector is operated by defining an exposure time, for which the detector system will measure before performing one readout operation to the control computer. This was set between 16 and 512 µs in the fibre location demonstrated here. During each exposure there are many laser pulses (at a rate of 20 to 80 MHz), the system only records the first photon to arrive on each pixel, and is unable to record another count until a readout operation has occurred (at the end of the exposure). To obtain undistorted TCSPC data, it is critical to operate in a regime where the detector does not have a high probability of photon detection for each laser pulse, otherwise photon pileup will occur where the TCSPC histogram becomes distorted to over represent early arriving photons as they are preferentially detected. In the system presented here it may be acceptable to count one photon per exposure (as this includes many thousands of laser pulses) without photon pileup, however in this work the exposure time was set to be below this level. The exception is Fig. 6(a) & (b), where some saturation may occur, as settings were chosen to match those of Fig. 6(c) & (d). Observation of less than 1 count per exposure ensures that the system is operated well away from the photon pileup regime. Furthermore, counting a photon on every pixel on every exposure distorts the relative amplitude of the signals on neighbouring pixels as pixels are saturated at 1 count per exposure. This would not be critical here, as only the timing information is key. However, exposure settings were chosen with the above considerations in mind depending on signal amplitudes in each of the experiments described (see Table 1). It should be noted that if a short exposure is set (<500 µs) the system spends idle time waiting for data readout in between exposures. Higher readout rates now available would shorten total measurement times in these cases.

Each pixel of the detector array has an independent TDC circuit, causing some variation in TDC bin duration on a per pixel basis. This causes cumulative errors in photon arrival time in later TDC bins. While this can be calibrated, in this work delay lines ensured that the photon detection occurred in early TDC bins to minimise this effect.

**Table 1, Experimental settings for data.**

|  | Field of view (cm) | Laser power (mW) | Pulse repetition rate (MHz) | Exposure time per frame (µs) | Number of frames | Total exposure time (s) | **Equivalent total exposure time at 7.4 mW (s)** |
|---|---|---|---|---|---|---|---|
| Fig 2, fibre: | 5 | 0.35 | 40 | 4096 | 56k | 230 | **11** |
| Fig 3, milk: | 15 | 0.18 | 20 | 512 | 220k | 110 | **2.7** |
| Fig 4, |  |  |  |  |  |  |  |
| avian model: | 15 | 0.10 | 20 | 128 | 110k | 14[a] | **0.19** |
| ovine lung with rib cage: | 15 | 0.10 | 20 | 512 | 1300k | 660 | **9.0** |
| Fig 5, ventilated ovine lung |  |  |  |  |  |  |  |
| lights on: | 30 | 6.8 | 80 | 16 | 45k | 0.72[a] | **0.66** |
| lights off: | 30 | 6.8 | 80 | 16 | 29k | 0.46[a] | **0.43** |
| Fig 6, |  |  |  |  |  |  |  |
| human hand: | 30 | 7.4 | 80 | 512 | 34k | 17 | **17** |
| human torso: | 30 | 7.4 | 80 | 512 | 34k | 17 | **17** |

[a] the experimental measurement time was longer than stated here due to detector readout overheads present when a short exposure time per frame is set

## 5.4 Data correction

Manufacturing variability and defects cause the MegaFrame detector to have a slightly non uniform response across the array. The most dominant effect is variability in Dark Count Rate (DCR) when no light is incident on the detector. In extreme cases, pixels can have a sufficiently high DCR that they are unable to measure incident photons in a useful way. Minimal processing of the detector response is implemented in the data presented here, however the DCR is measured for each pixel response from the region of the TCSPC histogram before any light arrives at the detector, and this value is subtracted from the entire trace.

In cases where the DCR is observed to be higher than a threshold (dependent on set parameters) the pixel is designated to be a 'screamer' and all data for this pixel is discarded. To avoid holes in the image, data is interpolated using an average of the 4 nearest neighbouring pixels.

Photon counting statistics and variability of the detector array timing properties cause some noise in the time axis for each pixel reducing image quality in the simple processing applied here. A 5 point adjacent averaging on the time axis (equivalent to 250 ps, approximately equal to the detector jitter) is applied, chosen to be consistent with the sharpest timing features observed to avoid accidental modification of timing statistics. A variability in the detection efficiency across the array is also observed, but this only affects the observed amplitude of the signals by a small factor, the principle described here is dependent on the timing information rather than exact magnitude of photon arrivals, so no correction is required.

*5.5 Models*

The coil of fibre in Fig. 2 was commercial 50 µm core silica glass without any external tubing, with laser light coupled via butt coupling to the fibre laser output. The coiled fibre was held to a board with a piece of transparent tape in the bottom left of the images (causing some increase in scattered output in this region). Field of view was 5 cm.

For Fig. 3, undiluted semi-skimmed milk (an emulsified colloid of ~1.5% liquid butterfat globules dispersed within a water-based solution) was contained within a tank dimensions ~30 cm x 20 cm x 20 cm (WxHxD). Optic fibre was positioned half way through this depth (~10 cm). A preliminary step of filling the tank with water and focussing the camera system confirmed good optical alignment and fibre location prior to replacing the water with milk. The lens used in this test provided a 15 cm field of view.

Fig. 4 data was taken with a horizontal separation between tissue models and the camera in an alternate lab bench setup with 15 cm field of view, allowing fibre probe insertion to be aided by gravity. Fig. 5 and Fig. 6 had a 30 cm field of view with the camera mounted on a standard tripod. The ventilated perfused lung model described elsewhere [6] was navigated by an experienced bronchoscope user to position the fibre probe in controlled lung segments. The lungs were retrieved from ewes destined for cull and were euthanized under Schedule 1 of Animals (Scientific Procedures) Act 1986.

*5.6 Accompanying visualisation descriptions*

Supplementary files provide experimental data, processed as described above, to produce video of light propagation through samples. Individual frames of this data are shown in the figures above. Where relevant, the earliest photon arrivals are highlighted in the video data. For close examination, the viewer is recommended to use a player that allows controlled frame by frame advancement through the video to view the few frames in which the early photons arrive before later scattered light. For closer examination, processed and raw data files are provided (see Data Availability). In all cases, the histogram data shown in the accompanying figures were used to examine photon arrival statistics for confirmation of earliest arriving photon locations.

### Visualisation 1

Compilation of subsequent visualisations, provided for convenience. Experimental arrangement of Fig. 1 is shown, followed by the visualisations listed below linked to the subsequent figures. See descriptions below for details.

Also available from http://dx.doi.org/10.7488/ds/2108.

[Visualisation 2](#)

Accompanying Fig. 2. Observation of 'light in flight' through an optical fibre (deliberately defocussed to increase spatial width of pulse for visibility). The laser pulse is seen to enter the field of view, guided in the fibre, from the top of the frame and circulate clockwise 6 times before exiting bottom right. Reflection from the fibre end causes a pulse of light to return shortly after. At the bottom left of the frame, the coil is fibre is held in position with adhesive tape, causing increased scattering at this location.

[Visualisation 3](#)

Accompanying Fig. 3. Laser pulse scattering through a uniform colloidal suspension (milk). Evidence of light leakage from the fibre is seen at the top right of the frame, followed by emission from the fibre tip (highlighted) within the medium. Photon detection increases spatially due to scattering as time progresses. The scattered photon cloud is seen to progress towards the bottom left of the frame, in the direction the fibre was orientated.

[Visualisation 4](#)

Accompanying Fig. 4. Fibre location in tissue models, a whole avian (chicken) model and an ovine lung with rib cage. Avian model: Evidence of light leakage from the fibre is seen at the bottom right of the frame, followed by emission from the fibre tip (highlighted) inside the avian model. As time progresses scattered light is observed to progress through the complex tissue structures and cavities, illuminating spatially distant structures, with decreasing light remaining at the source. Ovine lung and rib cage model: light leakage from the fibre is seen at the top right of the frame, followed by emission from the fibre tip (highlighted), deep inside the lung. Scattered light progresses, with greatest amplitude in the orientation of the fibre, through the tissue structures towards the bottom left of the frame.

[Visualisation 5](#)

Accompanying Fig. 5. Fibre location in a ventilated and perfused ovine lung model with the camera on a tripod over the 'patient', in ambient light conditions (fluorescent room lights) and in the 'dark'. Fibre is delivered through a bronchoscope, positioned by a skilled operator. In both cases, light is initially observed from the fibre tip (highlighted), then light scattered through the spatially separated tissue structures dominates the image. Notable complex structure is observed, likely due to light passing through airways with less attenuation than denser tissue. Between demonstrations the fibre is repositioned to a neighbouring lung segment, resulting in a shift in early photon emission location and scattering is observed through different complex structures.

[Visualisation 6](#)

Accompanying Fig. 6. Fibre location through the human body, a hand and a torso (below ribs, imaged from the front of body with the fibre placed against the back). For the hand, lesser scattered light is observed from the fibre tip (highlighted), followed by greater scattered light with increased spatial extent. Some evidence of the hand outline is observed. In the case of the torso, in a few image frames early arriving light from the fibre tip is observed (highlighted) followed by further scattered light with large spatial extent. Scattered light appears to progress to the right of the frame and to the bottom of the frame, highlighting unknown structures with preferential transmission properties away from the fibre tip.

**Funding**

The authors acknowledge the UK Engineering and Physical Sciences Research Council (EP/K03197X/1) for funding this work.

**Disclosures**

The authors declare that they have no competing financial interests. UK Patent application (number 1611819.2) pertaining to the optical endomicroscope location method has been filed.

**Acknowledgments**

We thank ST Microelectronics, Imaging Division, Edinburgh for their support in the manufacture of the Megaframe chip. The Megaframe project has been supported by the European Community within the Sixth Framework Programme IST FET Open. We thank Nikola Krstajić for providing CMOS SPAD support and providing the front end Labview software interfacing to Megaframe.

**Author Contributions**

M.G.T. developed the method and instrumentation for fibre optical endomicroscope location, performed the experiments, processed the data and drafted the manuscript. T.H.C., K.D., M.B., R.R.T. and R.K.H contributed to manuscript draft. T.R.C. performed bronchoscopy and fibre probe delivery in the ovine lung model, T.R.C., T.H.C., B.M. and K.D. developed the ex-vivo ventilated ovine lung model and aided with tissue samples. R.K.H. designed the CMOS SPAD pixel architecture. M.G.T., K.D. and R.R.T. conceived the concept, R.R.T. led the project. All authors contributed to the discussion of results and the project motivation.